\documentclass[graybox]{svmult}
\usepackage[utf8]{inputenc}
\usepackage{type1cm}        
\usepackage{makeidx}         
\usepackage{graphicx}        
\usepackage{multicol}        
\usepackage[bottom]{footmisc}

\usepackage{mathtools}

\usepackage{amsmath}
\usepackage{amssymb,mathrsfs}
\usepackage{mathtools}
\usepackage{verbatim}

\usepackage[unicode,psdextra]{hyperref}
\usepackage[sort&compress,capitalize,nameinlink]{cleveref}
\usepackage{enumerate}

\DeclareMathOperator{\tr}{Tr}

\DeclareMathOperator{\BCS}{BCS}

\DeclareMathOperator{\GP}{{GP}}

\DeclareMathOperator{\h}{\mathfrak{h}}

\DeclareMathOperator{\EBCS}{\mathcal{E}^{\BCS}_{\mu}[\Gamma]}
\newcommand{\EGP}{ \mathcal{E}_{D}^{\GP}(\p)}
\DeclareMathOperator{\tg}{\tr\h\gamma}

\DeclareMathOperator{\p}{\psi}
\DeclareMathOperator{\A}{\alpha}
\DeclareMathOperator{\Ap}{\A_{\p}}
\DeclareMathOperator{\Ao}{\alpha_{0}}

\newcommand{\R}{\mathbb{R}}

\newcommand{\lf}{\left}
\newcommand{\ri}{\right}

\newcommand{\braketr}[2]{\lf\langle #1\lf|#2\ri. \ri\rangle}

\newcommand{\mean}[1]{\lf\langle #1 \ri\rangle}
\newcommand{\meanlrlr}[3]{\lf\langle #1\lf|#2\ri|#3\ri\rangle}

\newcommand{\beq}{\begin{equation}}
\newcommand{\eeq}{\end{equation}}
\newcommand{\bdm}{\begin{displaymath}}
\newcommand{\edm}{\end{displaymath}}
\newcommand{\bdn}{\begin{eqnarray}}
\newcommand{\edn}{\end{eqnarray}}
\newcommand{\bay}{\begin{array}{c}}
\newcommand{\eay}{\end{array}}
\newcommand{\ben}{\begin{enumerate}}
\newcommand{\een}{\end{enumerate}}
\newcommand{\beqn}{\begin{eqnarray}}
\newcommand{\eeqn}{\end{eqnarray}}
\newcommand{\bml}[1]{\begin{multline} #1 \end{multline}}
\newcommand{\bmln}[1]{\begin{multline*} #1 \end{multline*}}

\renewcommand{\E}{\mathcal{E}}
\renewcommand{\leq}{\leqslant}
\renewcommand{\geq}{\geqslant}

\newtheorem{ass}{Assumption}

\crefname{ass}{Assumption}{Assumptions}
\crefname{section}{\textsection}{\textsection}

\title*{Derivation of the {Gross-Pitaevskii} Theory for Interacting Fermions in a Trap}
\titlerunning{Derivation of GP Theory for Interacting Fermions in a Trap}
\author{Andrea Calignano and Michele Correggi}
\institute{Dipartimento di Matematica, Politecnico di Milano, P.zza Leonardo da Vinci, 32, 20133, Milano, Italy}
\date{\today}

\begin{document}

\maketitle

\abstract{We study a dilute gas of interacting fermions at temperature $T=0$ and chemical 
potential $\mu \in \mathbb{R}$. The particles are trapped by an external potential, and they interact via a microscopic attractive 
two-body potential with a two-body bound state. We prove the emergence of the macroscopic 
{Gross-Pitaevskii} theory as first-order contribution to the BCS energy 
functional in the regime of vanishing micro-to-macro scale parameter.}


\section{Introduction}
\label{sec: intro}

The low-temperature behavior of interacting fermions has been widely studied in the physics literature {(see, e.g, the monographs \cite{L:book,T:book})}, in order to understand phenomena as the occurrence of superconductivity in materials, i.e., a sudden drop of resistivity below a certain critical temperature. A microscopic model for such a phenomenon was proposed in the `50s in \cite{BCS:paper} by  J. Bardeen, L. Cooper and R. Schrieffer, and it is nowadays very well known as the BCS theory: the presence of an attraction between the fermions may be responsible for the formation of (weakly) bound pairs (Cooper pairs) of fermions with opposite spin; such pairs behave in all respect as charged bosons and as such they undergo Bose-Einstein condensation below a certain critical temperature. The emergence of this collective behavior of Cooper pairs is the signature of the {occurrence of superconductivity in} the material{, and it can be {understood starting from} the minimization of the free energy of the system given by the {\it BCS energy functional}} depending on the two-particle reduced density matrix.

Few years before the appearance of the BCS description of superconductivity, a much more phenomenological macroscopic explanation was provided in \cite{GL:paper} by V.L. Ginzburg and L.D. Landau. In the GL theory the superconducting features of the sample are encoded in an order parameter $ \psi $, i.e., a complex wave function minimizing a suitable energy functional, which is supposed to approximate the free energy of the system {(see \cite{AGia:paper, C:paper, CG1:paper, CG2:paper, KP:paper} and references therein for some recent mathematical results)}. The connection between the two models was heuristically investigated in \cite{G:paper}, but only much more recently a rigorous derivation of GL theory from the BCS model was obtained in \cite{FHSS:paper} (see also the related papers \cite{DHS:paper, FHL:paper,FHSS1:paper, HSc:paper, HS1:paper}): it is shown that, in a translational invariant system in presence of slowly varying external potentials and close to the critical temperature for the superconductivity transition, the leading order of the BCS {ground state} energy is given by the {minimum} of the GL functional, provided the attraction admits at least a bound state and in {the} limit of zero ratio between the microscopic scale of the interaction and the macroscopic {size} of the sample. The zero-temperature analogue of the same result for a fermionic system in a bounded domain was {successively} obtained in \cite{FLS:paper}, while a similar question for the Bogolubov-Hartree-Fock functional, i.e., the BCS energy functional with the addition of direct and exchange terms, was studied in \cite{BHS:paper}.

The setting we consider here is quite close to the one addressed in \cite{FLS:paper,HS1:paper}, i.e., we study the zero-temperature behavior of a gas of interacting fermions, but, unlike the previous references, here we assume the presence of a {\it confining external potential}. The particles interact via a two-body attraction, which is strong enough to bind two particles together. Naively, one may think that the fermions at low temperature would arrange in bounded pairs, so forming a bosonic gas, which then undergoes BE condensation. However, as in \cite{BHS:paper, FLS:paper, HS1:paper}, one observes that the possibility to form a two-body bound state is in fact enough to generate the superconductivity transition, even though the gas does not exactly arrange in two-particle bound pairs.  

Let us describe the setting more precisely: we set the length scale of the trap to be $ 1 $, while the microscopic interaction varies on a scale $ h \ll 1 $. The parameter $ h $ thus describes the ratio between the micro- and macroscopic scales and we study the limit $ h \to 0 $ of the ground state energy of the BCS energy functional and of any corresponding minimizer.  We do not fix the number of particles a priori, but we study the grand-canonical problem in presence of a chemical potential $ \mu $.

We stress that the physical setting we are considering is not the typical one of BCS theory in which the formation of Cooper pairs occurs on a scale much larger than the mean interparticle distance. On the {contrary}, here, the size of bounded pairs is of order $ h $ and it is therefore much smaller than the mean distance travelled by fermions, which, as we are going to see, is of order $ h^{1/3} $ (the density of particles if of order $ h^{-1} $). There is however a physical regime in which this setting becomes meaningful, namely the BEC/BCS crossover region {(see \cite{HS1:paper})}, where for certain values of the two-particle scatting length, the picture is very close to the one considered here. Note also that, as a gas made of almost bosonic pairs, the system is dilute {(see also next \cref{rem: diluteness} and the analogous discussions in \cite{DSY:paper, LSY:paper}),} because the density times the microscopic volume where the interaction acts non-trivially is of order $ h^{-1} \cdot h^3 = h^2 \ll 1 $.
 
\begin{acknowledgement}
	The authors thank \textsc{C. Hainzl} for useful comments and remarks about the physical interpretation of the model. The support of Istituto Nazionale di Alta Matematica (INdAM) ``F. Severi'' via the intensive period \href{https://sites.google.com/view/iqm22}{``INdAM Quantum Meetings (IQM22)''} is also acknowledged. 
\end{acknowledgement}

\subsection{BCS theory of superconductivity}
In the BCS model all the information about the state of the system is encoded in two variables: the reduced one-particle density matrix $\gamma$ and the pairing density matrix $\alpha$. Hence, the system is fully described by an operator
\begin{equation}
\Gamma =
\begin{pmatrix}
    \gamma & \alpha \\
    \bar{\alpha} & 1 - \overline{\gamma}
\end{pmatrix}, \qquad 0\leq \Gamma\leq 1,
\end{equation}
acting on $L^{2}(\mathbb{R}^{3})\oplus L^{2}(\mathbb{R}^{3})$. The bar denotes complex conjugation, i.e., the integral kernels of the operators $\overline{\gamma}, \overline{\alpha}$ are $\overline{\gamma(x,y)}$ and $\overline{\alpha(x,y)}$, respectively. For a given $\BCS$ state $\Gamma$, the $\BCS$ functional at $T=0$ in macroscopic units is given by
\begin{equation}
    \EBCS \vcentcolon= \tg+\int_{\mathbb{R}^{6}}dx dy \: V\lf(\tfrac{x-y}{h}\ri)\lvert\alpha(x,y)\rvert^2 ,
\end{equation}
where the one-body operator $\h=-h^2\Delta+h^2W-\mu$ describes the energy of non-interacting electrons at chemical potential $\mu<0$. The BCS ground state energy is
\begin{equation}
	E^{\BCS}_{\mu} : = \inf_{0 \leq \Gamma \leq 1} \EBCS.
\end{equation}

\begin{ass}[Existence of a {ground} state]
\label{assumption:V}
We assume that $V$ is real, radially symmetric, locally integrable and bounded from below. Moreover, the two-particle operator $-\Delta + V$ is assumed {to admit a normalized ground} state $\alpha_{0} \in L^2(\R^6) $ with corresponding energy $-E_{0}, \ E_{0}>0$, which in particular implies that the negative part of $ V $ is non-zero.
\end{ass}

\begin{ass}[Spectral gap]
\label{hypothesis:spectralgap}
Let $\alpha_{0}$ be the ground state as in \cref{assumption:V} above. We assume that $ \exists g>0 $ and $ 0 < \varepsilon < 1$, such that
\begin{equation}
    P_{\alpha_0}^{\perp} \lf[ -(1-\varepsilon)\Delta+V+E_{0} \ri] P_{\alpha_0}^{\perp} \geq g P_{\alpha_0}^{\perp} 
\end{equation}
where $ P_{\alpha_0}^{\perp} $ stands for the projector onto the orthogonal complement of $\alpha_{0}$.
\end{ass}

{\begin{ass}[Trapping potential]
\label{assumption:W}
We also assume that $W\in C^{1}(\R^3)$ is positive and  there exist $ 0 < \beta, c_1, c_2 < + \infty $ such that
\begin{equation}
        		 \begin{cases}
        		 	c_1 |x|^\beta \leq W(x) \leq c_2  |x|^{\beta},	\\
        		 	\lf| \nabla W (x) \ri| \leq c_2 \beta |x|^{\beta-1},
			\end{cases}
			\qquad		\mbox{for } |x| \geq 1.
\end{equation}
\end{ass}}

We stress that for the class of attractive potentials in \cref{assumption:V}, one can deduce by standard Agmon estimates (see, e.g.,\cite{AG:paper}) the exponential decay of the bound state wave function $\Ao$: there exists $b>0$ such that
\begin{equation}
\label{property:decay}
    \int_{\R^{3}} dx \: \lvert \Ao(x) \rvert^{2} e^{2b x} < +\infty.
\end{equation}
Note also that \cref{assumption:W} allows to Taylor expand
\begin{equation}
\label{Taylor}
    W(\eta+\xi/2) = W(\eta) + \tfrac{\xi}{2} \cdot \nabla W(\zeta),
\end{equation}
with the variable $\zeta$ belonging to $(\eta,\eta+\xi/2)$. 
A special case of a potential satisfying \cref{assumption:W} is obviously given by the harmonic potential. In this case, the two-body Hamiltonian perfectly decouples in relative and centre-of-mass coordinates, which allows to get rid of several error terms {in the discussion below}.

The condition $0\leq \Gamma\leq 1$, which is often call admissibility of $ \Gamma $, implies that the operator $\gamma$ is hermitian, i.e. $\gamma(x,y)=\overline{\gamma(y,x)}$ and that $\alpha$ is such that {$\overline{\alpha}=\alpha^{\dagger}$}. Furthermore, the operators $\gamma,\alpha:L^{2}(\R^{3}) \rightarrow L^{2}(\R^{3})$ have a specific physical meaning (see, e.g., \cite{BLS:paper} for a formal derivation): given a many-body fermionic state $ \Psi $, we have
\begin{equation}
    \gamma(x,y) = \mean{a_{x}^{\dagger}a_{y}}_{\Psi},	\qquad    \alpha(x,y)  =\mean{a_{x}a_{y}}_{\Psi}
\end{equation}
i.e., they represent the one-particle density matrix of the system and the wave function of a Cooper pair, respectively. Here $a_{x}^{\dagger},a_{x}$ are the fermionic creation and annihilation operators. In fact, in absence of any pairing between the fermions, the system is in the so-called {\it normal state}, which is characterized by a trivial off-diagonal component, i.e., $ \alpha \equiv 0 $. The emergence of superconductivity is then associated to a non-trivial $ \alpha $.

\subsection{{Gross-Pitaevskii} theory}
For any $D \in \R$, the {Gross-Pitaevskii (GP)}  energy functional is defined as
\begin{equation}
\label{equation:GPfunctional}
    \EGP \vcentcolon= \int_{\R^{3}} d\eta \ \lf\{ \tfrac{1}{4}\lvert \nabla\psi  \rvert^{2} + (W(\eta)-D) \lvert \psi \rvert^{2} + g_{\BCS}\lvert \psi  \rvert^{4} \ri\},
\end{equation}
where the coefficient $g_{\BCS}>0$ represents the interaction strength among different pairs, and whose expression in terms of the microscopic quantities is provided in \cref{thm:energy}. The {GP} energy can be proven to be bounded from below for any positive $ g_{\BCS} $ {(see \cref{corollary: bounded from below})}. We denote then the {GP} ground state energy by
\begin{equation}
    E^{\GP}_{D}\vcentcolon=\inf_{\psi \in \mathscr{D}^{\GP}} \EGP,
\end{equation}
where $\mathscr{D}^{\GP}=\{ \p \in H^{1}(\mathbb{R}^{3}) | W\lvert \p \rvert^{2} \in L^{1}(\mathbb{R}^{3}) \}$ is the natural minimization domain for \eqref{equation:GPfunctional}. We denote by $ \psi_{*} $ the corresponding minimizer, which can be shown to be unique up to choice of the phase by strict convexity of the functional in $ |\psi|^2 $.

{We point out here that, mathematically speaking, the GP functional introduced above may as well be named {\it Ginzburg-Landau} functional, although the energy does not look exactly as the usual GL energy, which in a homogeneous sample would read
\beq
	\label{eq: GL energy}
	\mathcal{E}^{\mathrm{GL}}[\phi] = \int_{\R^3} d\eta \ \lf\{ \tfrac{1}{4} \lf| \nabla\phi  \ri|^{2} + \tilde{g}_{\BCS} \lf( 1 - \lf| \phi  \ri|^{2} \ri)^2 \ri\}.
\eeq	
However, it is possible (see below and the discussion in \cite[Sect. 1]{CPRY:paper}) to reduce the minimization of \eqref{equation:GPfunctional} to the one of a functional very close to \eqref{eq: GL energy} (in fact, its inhomogeneous counterpart).} 

{Notice that the GP wave function $ \psi $ is not normalized in $ L^2 $ since we are performing the energy minimization in the grand canonical setting, and therefore we may think that  $ \lf\| \psi \ri\|_2 $ is determined by the value of the chemical potential $ \mu $. Let us denote by $ N $ such a quantity, i.e., $ N : = \lf\| \psi_{*} \ri\|_2^2 $, and let $ f_0 $ be the positive minimizer of the GP energy
\bdm
	\widetilde{\mathcal{E}}^{\mathrm{GP}}[f] = \int_{\R^{3}} d\eta \ \lf\{ \tfrac{1}{4}\lvert \nabla f  \rvert^{2} + W(\eta) \lvert f \rvert^{2} + g_{\BCS}N \lvert f  \rvert^{4} \ri\},
\edm
with $ L^2$ norm set equal to $ 1 $. Such a minimizer satisfies the variational equation 
\bdm
	- \tfrac{1}{4} \Delta f_0 + W f_0 + 2 g_{\BCS}N f_0^3 = \mu_0 f_0,
\edm
for a chemical potential $ \mu_0 =  \widetilde{{E}}^{\mathrm{GP}} + g N \lf\| f_0 \ri\|_4^4 $, where we have set  $ \widetilde{E}^{\mathrm{GP}} : = \inf_{\lf\| \psi \ri\|_2 = 1} \widetilde{\mathcal{E}}^{\mathrm{GP}}[\psi] $. With the splitting $ \psi_{*} = : \sqrt{N} f_0 \phi_* $ and exploiting the variational equation for $ f_0 $, one gets 
\bdm
	E^{\mathrm{GP}}_D = N \lf\{ \widetilde{E}^{\mathrm{GP}} - D + \widetilde{\mathcal{E}}^{\mathrm{GL}}[\phi_*] \ri\},
\edm
where the last term is a weighted Ginzburg-Laudau functional explicitly given by
\beq
	\widetilde{\mathcal{E}}^{\mathrm{GL}}[\phi] = \int_{\R^3} d \eta \ f_0^2 \lf\{ \tfrac{1}{4} \lf| \nabla\phi  \ri|^{2} + \tilde{g}_{\BCS} N f_0^2 \lf( 1 - \lf| \phi  \ri|^{2} \ri)^2 \ri\},
\eeq
and $ \phi_* $ its minimizer.}

{This makes apparent the connection between the GP and GL functionals, so that, from this point of view, both names are mathematically equivalent to identify \eqref{equation:GPfunctional}. There is however a strong physical motivation (see also \cite{HS1:paper}) for the choice we made, namely the fact that the physical regime we are investigating is a BEC one: as described in \cref{sec: intro}, the mechanism behind the emergence of a collective behavior in the low-temperature Fermi gas considered here is not the usual BCS pairing phenomenon, but rather a condensation of fermionic pairs playing the role of bosonic molecules. The pairs have indeed a size of order $ h \ll 1 $ which is much smaller that the typical distance between the fermionic particles of order of the trap length scale $ O(1) $.}

\section{Main Results}
This section contains our main results about the semiclassical expansion of the $\BCS$ energy. 

\begin{theorem}[BCS energy]
\label{thm:energy}
Let $\mu=-E_{0}+Dh^{2}$, for some $D\in\R$ and let \cref{assumption:V,hypothesis:spectralgap,assumption:W} be satisfied. Then,
\begin{equation}
    E_{\mu}^{\BCS} = h E_{D}^{\GP} + O(h^{2}),
\end{equation}
as $h \to 0$, where
    	\begin{equation}
       	g_{\BCS}\vcentcolon=  (2\pi)^{3}\int_{\R^{3}} dp \: (p^{2}+E_{0})\lvert \hat{\alpha}_0(p)\rvert^{4}.
	\end{equation}
	Moreover, for any approximate ground state $ \Gamma $ of the BCS functional, i.e., such that 
\begin{equation}
     \mathcal{E}_{\mu}^{\BCS}[\Gamma] \leq E_{\mu}^{\BCS} + \varepsilon h, \qquad 0<\varepsilon {< + \infty},
\end{equation}
its off-diagonal element $\alpha$ can be decomposed as 
\begin{equation}
	\label{eq: alpha}
    \alpha(x,y) = h^{-2}\psi\lf(\tfrac{x+y}{2}\ri)\alpha_{0}\lf(\tfrac{x-y}{h}\ri)+r(x,y), 
\end{equation}
where $\p \in \mathscr{D}^{\GP} $ satisfies $\mathcal{E}_{D}^{\GP}(\p) \leq E_{D}^{\GP} + \varepsilon {+ o(1)}$, {$\alpha_{0}$ is the ground state of the two-particle operator} and the correction $r$ is small in the following sense:
\begin{equation}
    \lVert r \rVert_{L^{2}}^{2} = O(h) , \qquad \lVert \nabla r \rVert_{L^{2}}^{2} {+ \lf\| W |r|^2 \ri\|_{L^1}} = O(h^{-1}).
\end{equation}
\end{theorem}

\begin{remark}[Diluteness]
	\label{rem: diluteness}
	The expansion \eqref{eq: alpha} together with the heuristics $ \gamma \simeq \alpha \overline{\alpha} $ (see \cref{section:LB}) suggests that the density of the gas in our setting is proportional\footnote{In fact, it may be possible to prove a weak version of such a statement as in \cite[Proposition 1.11]{FLS:paper} using the Griffith's argument, i.e., variation w.r.t. to the external potential. However, we omit this discussion here for the sake of brevity.}  to $ h^{-1} \lf| \psi \ri|^2 $, i.e., the total number of particle is of order $ h^{-1} $. This vindicates the statement about the diluteness of the system since the range of the two-body interaction is $ \propto h $ and therefore the diluteness parameter $ h^{-1} h^3 = h^2 \ll 1 $ is small. 
\end{remark}

{\begin{remark}[Properties of $ \alpha_0 $]
	\label{rem: alpha0}
Note that by the estimate \eqref{property:decay}, $ \alpha_0 \in L^{1} \cup H^1(\R^3) $, which guarantees that $ \hat{\alpha}_0 \in L^{\infty}(\R^3) $, so that $ \hat{\alpha}_0 \in L^p (\R^3) $ for any $ p \geq 2 $ and $g_{\BCS} $ is a finite quantity.
\end{remark}}

Whether the systems is superconducting in the asymptotic regime $ h \to 0 $ thus depends on the fact that the {GP} {wave function} $ \psi $ is non-trivial. For the {GP} minimizer this depends on the value of the coefficient $ D $, which in turn is determined by the chemical potential $ \mu $. In fact, one can infer \cite{FLS:paper, HS1:paper} from the properties of the function
\begin{equation}
    \mu \mapsto E_{\mu}^{\BCS},
\end{equation}
which is continuous, concave, and monotone decreasing, that there exists a unique critical value $\mu_{c}(h)$ such that below $ \mu_c $ superconductivity is present and above it the system is in the normal state. The exact definition of $  \mu_{c}(h) $ is the following:
\begin{equation}
    \mu_{c}(h) \vcentcolon= \inf \lf\{ \mu < 0 \ \big| \ E_{\mu}^{\BCS} < 0 \ri\},
\end{equation}
i.e., it marks the threshold of the transition from a zero ground state energy (normal state) to a strictly negative one. 

\begin{theorem}[Critical chemical potential]
\label{thm:chemicalpot}
Under the assumptions of \cref{thm:energy}, the critical chemical potential at which the superconductivity phase transition takes place is \begin{equation}
    \mu_{c}(h)=-E_{0}+E_{W}h^{2} + o(h^2),    
\end{equation}
as $h\rightarrow 0$, where $ E_{W} $ is the ground state energy of the one-particle operator $ -\frac{1}{4}\Delta + W $.
\end{theorem}

\section{Proofs}

The key ingredient to prove \cref{thm:energy} and \cref{thm:chemicalpot} is given by the following \cref{theorem:MAING}, which provides the link between the BCS and GL functionals.

\begin{proposition}[BCS and {GP} functionals]
\label{theorem:MAING}
Let $\mu=-E_{0}+Dh^{2}$, $D\in \mathbb{R}$. Then,
\begin{enumerate}[(a)]
    \item {\it Upper bound}: for any $\psi \in \mathscr{D}^{\GP}$ there exists an admissible state $\Gamma_{\psi}$ such that 
    \begin{equation}
    		\label{eq: upper bound}
        \mathcal{E}^{\BCS}_{\mu}[\Gamma_{\psi}] \leq h \EGP + C h^2 \lf[ 1 + \lf(\max\lf\{\EGP, 0 \ri\}\ri)^2 \ri].
    \end{equation}
    
    \item	{\it Lower bound}: let $\Gamma$ be an admissible $\BCS$ state such that $\mathcal{E}^{\BCS}_\mu[\Gamma] \leq C_{\Gamma} h$. Then, there exists $\psi \in \mathscr{D}^{\GP}(\mathbb{R}^{3})$ such that 
    \begin{equation}
    		\label{eq: lower bound}
        \mathcal{E}^{\BCS}_{\mu}[\Gamma] \geq h \EGP - C h^2.
    \end{equation}
    Furthermore, there exists a function $r$ such that the following decomposition holds: 
    \begin{equation}
    \label{equation:decomposition}
        \alpha(x,y) = h^{-2}\psi\lf(\tfrac{x+y}{2}\ri)\alpha_{0}\lf(\tfrac{x-y}{h}\ri)+r(x,y).
    \end{equation}
    where the remainder $ r $ satisfies the bounds
	\begin{equation}
    \label{equation:bounds on xi}
	    \lVert r \rVert_{L^{2}}^{2} \leq Ch, \qquad \meanlrlr{r}{-\Delta + W}{r}_{L^{2}}  \leq C h^{-1}.
\end{equation}
\end{enumerate}
\end{proposition}

Let us then assume that \cref{theorem:MAING} holds and prove \cref{thm:energy} and \cref{thm:chemicalpot}. {The proof of \cref{theorem:MAING} will be given in next \cref{section:UB} and \cref{section:LB} by separately addressing points $ (a) $ and $ (b) $ of the statement.}

\begin{proof}[\cref{thm:energy}]
	To prove the upper bound, we use the admissible trial state $\Gamma_{\psi_{*}}$, where we recall that $ \psi_* $ stands for the minimizer of the {GP} functional. We then obtain by \eqref{eq: upper bound}
    \begin{equation}
        E_{\mu}^{\BCS} \leq \mathcal{E}_{\mu}^{\BCS}[\Gamma_{\psi_{*}}] = h \mathcal{E}_{D}^{\GP}(\psi_{*})+O(h^{2})= h E_{D}^{\GP}+O(h^{2}),
    \end{equation}
    since $ \mathcal{E}_{D}^{\GP}(\psi_{*}) = E^{\GP}_{D} \leq 0 $. {In addition to proving a sharp upper bound for the ground state energy, the estimate above also} yields the a priori bound $ \mathcal{E}_{\mu}^{\BCS}[\Gamma] \leq C h $ for any approximate minimizer $ \Gamma $ of the BCS energy. {Hence, the minimizer satisfies} \eqref{eq: lower bound}, so that we can deduce the estimate from below matching the upper bound, {along with} the decomposition of $ \alpha $ as in \eqref{equation:decomposition}.
\end{proof}

\begin{proof}[\cref{thm:chemicalpot}]
We start from the trivial observation that
\begin{equation}
	\label{eq: condition D}
	E_D^{\GP} < 0,	\quad \Longleftrightarrow \quad D > E_W,
\end{equation}
where we recall that $E_W $ is the ground state energy of $ - \frac{1}{4} \Delta + W $: indeed, if $ D > E_{W} $, it suffices to use $ \lambda \psi_W $, $ \lambda > 0 $, as a trial state for the {GP} energy, where $ \psi_W $ si the normalized ground state of $ - \frac{1}{4} \Delta + W $, to get
\begin{equation}
	E^{\GP}_D = \lambda (E_W - D) + g_{\BCS} \lambda^2 \lf\| \psi_W \ri\|_{L^4}^4 < 0,
\end{equation}
for $ \lambda $ small enough. On the other hand, if $ D \leq E_W $, the functional is trivially positive, since
\begin{equation}
	\label{eq: lower bound GL}
	\mathcal{E}^{\GP}_{D}(\p) \geq \lf(E_W - D\ri) \lf\| \psi \ri\|_{L^2}^2.
\end{equation}
Note also that $ \psi_* $ is non-trivial if and only if $ E_D^{\GP} < 0 $.

Next, we prove the upper bound $ \mu_c(h) \leq -E_{0}+E_{W}h^{2} + o(h^2) $ by showing that, if $\mu=-E_{0}+D h^{2} $, $ D > E_W $, then there exists an admissible $\BCS$ state such that
\begin{equation}
    \mathcal{E}_{\mu}^{\BCS}[\Gamma] < 0.
\end{equation}
By \cref{theorem:MAING}, for any $\p \in \mathscr{D}^{\GP}$, there exists $\Gamma_{\p}$ admissibile such that 
\begin{equation}
    h^{-1}\mathcal{E}_{\mu}^{\BCS}[\Gamma_{\p}]=\mathcal{E}^{\GP}_{D}(\p)+O(h).
\end{equation}
This in particular holds true for $ \psi = \psi_* $, so that
\begin{equation}
	E^{\BCS}_\mu \leq E^{\GP}_D h + O(h^2) < 0,	
\end{equation} 
if $ D > E_W $. 

Conversely, we now show that, if $ E^{\BCS}_\mu = 0 $ for a certain $ \mu = - E_0 + D h^2 $, then $ D \leq E_W $, so completing the proof. By \cref{thm:energy}, indeed, if $ E^{\BCS}_\mu = 0 $, then $ E^{\GP}_D = O(h) $ but the {GP} functional is independent of $ h $ and therefore $ E^{\GP}_D = 0 $, which in turn implies that $ D \leq E_W $ by \eqref{eq: condition D}.
\end{proof}

\subsection{{GP} functional}

We discuss some useful properties of the {GP} functional \eqref{equation:GPfunctional} and its minimization. We recall that we denote by $ E^{\GP} $ the infimum of \eqref{equation:GPfunctional} and by $ \psi_{*} $ any associated minimizer.

	\begin{proposition}[A priori bounds on $ \psi $]
		\label{pro: a priori psi}
		{There exists $ C < + \infty $, depending on $ g_{\BCS} > 0 $}, such that
		\beq
			\label{eq: a priori psi}
			\lf\| \nabla \psi \ri\|^2_{L^2} + \meanlrlr{\psi}{W}{\psi} + \lf\| \psi \ri\|_{L^4}^4  + \lf\| \psi \ri\|^2_{L^2}			
			\leq C \lf[ 1 + \max\lf\{\EGP, 0 \ri\} \ri]
		\eeq
  {for all $ \psi \in \mathscr{D}^{\GP} $}.
    \end{proposition}
	
	\begin{proof}
		We may assume that $ D \geq 0 $ otherwise the result is trivially obtained with $ C = {\max\lf\{|D|^{-1}, g_{\BCS}^{-1}, 4 \ri\}} $.
		The starting point is the inequality
		\bml{
			\meanlrlr{\psi}{-\tfrac{1}{4} \Delta + W}{\psi} + g_{\BCS} \lf\| \psi \ri\|_{L^4}^4 \leq D \lf\| \psi \ri\|_{L^2}^2 + \EGP \\
			\leq D \lf\| \psi \ri\|_2^2 + \max \lf\{ \EGP, 0 \ri\},
		}
		which allows to bound from above both the quantities on the l.h.s. in terms of the $ L^2 $ norm and the {GP} energy of $ \psi $. Next, we estimate for $ R $ large enough
		\bmln{
			\lf\| \psi \ri\|_{L^2}^2 \leq \int_{|x|\leq R} d x \: |\psi|^2 + R^{-\beta} \int_{|x|{>} R} d x \: |x|^{\beta} |\psi|^2	\\
			\leq \sqrt{\tfrac{4 \pi}{3}} R^{3/2} \lf\| \psi \ri\|_{L^4}^2 + C {R^{-\beta}}\meanlrlr{\psi}{W}{\psi}  \\
			 \leq C \lf[ R^{3/2} g_{\BCS}^{-1} \lf( D \lf\| \psi \ri\|_{L^2} + \sqrt{E} \ri) + R^{-\beta} \lf( {\sqrt{D}} \lf\| \psi \ri\|_{L^2}^2 + E \ri) \ri] 
		}
		where we have set $ E : = \max \lf\{ \EGP, 0 \ri\} $ for short. Hence, for $ R  > (CD)^{1/\beta} $, we get
		\bdm
			\lf( 1 - \tfrac{C D}{R^{\beta}} \ri) \lf\| \psi \ri\|_{L^2}^2 - C  R^{3/2}  D \lf\| \psi \ri\|_{L^2} \leq C  \lf( R^{3/2} g_{\BCS}^{-1}  \sqrt{E}  + R^{-\beta}  E \ri),
		\edm
		which implies
		\beq
			\lf\| \psi \ri\|_{L^2}^2 \leq \frac{C}{\lf(1 - \tfrac{C D}{R^{\beta}}\ri)^2} \lf[ \lf(1 - \tfrac{C D}{R^{\beta}}\ri) \lf( R^{3/2} g_{\BCS}^{-1}  \sqrt{E}  + R^{-\beta}  E\ri) + C^2  R^{3}  D^2 \ri]
		\eeq
		and thus the result.	
	\end{proof}
	
	\begin{corollary}[Boundedness from below of $ \EGP $]
		\label{corollary: bounded from below}
		For any $ g_{\BCS} > 0 $, there exists a finite constant $ C < + \infty $ such that
		\beq
			E^{\GP}_D \geq - C.
		\eeq
	\end{corollary}
	
	\begin{proof}
		{Again, $  E^{\GP}_D = 0 $, if $ D \leq 0 $, and there is nothing to prove, so let us assume that $ D > 0 $. In this case i}t suffices to observe that $ E^{\GP}_D \leq 0 $, which can be obtained by simply testing the {GP} energy on the trivial wave function $ \psi \equiv 0 $. Hence, \cref{pro: a priori psi} implies that $ \exists C < + \infty $ such that $ \lf\| \psi \ri\|_{L^2}^2 \leq C $ for any $ \psi $ with non-positive energy, which in turn yields the lower bound $ E^{\GP}_D \geq - C|D| $ and thus the result.	
	\end{proof}
	
	The existence of a minimizer $ \psi_* $ which is also unique up to gauge transformation can be deduced by standard methods in variational calculus, and any such a minimizer solves the variational equation
	\beq
		- \tfrac{1}{4} \Delta \psi_* + (W-D) \psi_* + 2 g_{\BCS} |\psi_*|^2 \psi_* = 0.
	\eeq
	Under \cref{assumption:W}, one can also show that $ \psi_* \in C^3 \cap L^{\infty} (\R^3) $ and it can be chosen strictly positive.

\subsection{Semiclassical estimates}

Before attacking the proof of \cref{theorem:MAING}, it is useful to state some technical but standard semiclassical bounds to be used in the rest of the paper.

\begin{proposition}[Semiclassical estimates]
\label{prop:semiclassics for ub}
Let $\mu=-E_{0}+h^{2}D$, $D\in \R$ and let 
\begin{equation}
\label{equation:alpha_p}
\Ap(x,y):=h^{-2}\p\lf(\tfrac{x+y}{2}\ri)\Ao\lf(\tfrac{x-y}{h}\ri), 
\end{equation}
for any $\p \in \mathscr{D}^{\GP} $. Then, the following estimates hold as $ h \to 0 $:
		\bml{
		\label{eq: semiclassical 1}
        \lf| \tr \h \A_{\p} \overline{\A_{\p}} + \int_{\mathbb{R}^{6}} dx dy \: V\lf(\tfrac{x-y}{h}\ri)\lvert\alpha_{\p}(x,y)\rvert^2 \ri. \\
        \lf. - h \meanlrlr{\psi}{- \tfrac{1}{4} \Delta + W - D}{\psi}_{L^2(\R^3)} \ri| \leq \mathrm{A}_{0} h^{2},
        }
    	where
		\beq
			\label{eq: A0}
			\mathrm{A}_0 = C \lf( \lf\lVert  {W} \lvert \p\rvert^{2} \ri \rVert_{L^1} + \lf\| \psi \ri\|_{L^2}^2 \ri);
		\eeq
    	\bml{
    	    \label{equation:quartic}
    		\lf| \tr \h \A_{\p}\overline{\A_{\p}}\A_{\p}\overline{\A_{\p}} - h g_{\BCS} \lVert\p\rVert^{4}_{L^{4}} \ri| \\
    		\leq  C h^{2} \lf[  \lVert \nabla\p\rVert^{4}_{L^{2}} + \lf\lVert W \lvert\p\rvert^{2} \ri\rVert_{L^{1}}^2  + \lf\| \psi \ri\|_{L^2}^4 + A_0 h \ri].
    }
\end{proposition}

Before discussing the proof of the above Proposition, it is convenient to state a technical result about the reduced density $ \alpha_{\psi} $ which is going to be used several times. In the following we will often use the center-of-mass coordinates
\beq
	\label{eq: center of mass}
	\eta : = \tfrac{1}{2} (x + y),	\qquad		\xi : = x - y,
\eeq
and use the notation
\beq
	 \tilde{\alpha}_{\psi}(\eta, \xi) := \alpha_{\psi}(x,y).
\eeq

\begin{lemma}
\label{proposition:estimates on alpha}
    Let $ \alpha_{\psi} $ be as \eqref{equation:alpha_p}. Then, for any $ n \in \mathbb{N} $ even,
    \beqn
            \label{equation:alphaphinorm}
            \lVert \alpha_{\p} \rVert_{\mathfrak{S}^{n}}^{n} & \leq &  C h^{n-3} \lVert \p \rVert^{n}_{L^{n}} \lf\lVert \lvert\hat{\alpha}_0\rvert \ri\rVert_{L^{n}}^{n},	\\ 
        \label{equation:nabla(x-y)}
            \lVert \nabla_\xi \tilde\alpha_{\p} \rVert_{{\mathfrak{S}}^{n}}^{n} &\leq& h^{-3}\lVert \p \rVert^{n}_{L^{n}} \lf\lVert \lf| \, \cdot \, \ri| \hat{\alpha}_0 \ri\rVert_{L^{n}}^{n},	
        \eeqn
	    where $\lVert \cdot \rVert_{\mathfrak{S}^{n}}$ stands for the Schatten norm of order $ n \in \mathbb{N} $.
\end{lemma}

\begin{proof}
	See \cite[Lemma 1]{BHS:paper}. The extension to any $ n \in \mathbb{N} $ is obtained by simply observing that, thanks to the monotonicity of Schatten norms, $ \lf\| \alpha_\psi \ri\|_{\mathfrak{S}^{\infty}} \leq \lf\| \alpha_\psi \ri\|_{\mathfrak{S}^{n}} $ for any $ n \in \mathbb{N} $, which allows to use \eqref{equation:alphaphinorm} and \eqref{equation:nabla(x-y)} repeatedly to extend the result to all natural numbers.
\end{proof}


We are now in position to present the proof of \cref{prop:semiclassics for ub}.

\begin{proof}[\cref{prop:semiclassics for ub}]
	Using the change to  center-of-mass and relative coordinates, one gets
    \bml{
    	\tr \h \A_{\p}\overline{\A_{\p}} + \int_{\mathbb{R}^{6}} dx dy \: V\lf(\tfrac{x-y}{h}\ri)\lvert\alpha_{\p}(x,y)\rvert^2 \\ 
    	=\meanlrlr{\tilde{\A}_{\p}}{-\tfrac{1}{4}h^{2}\Delta_{\eta} + h^{2}W(\eta+\xi/2) - h^{2}D}{\tilde{\A}_{\p}}_{L^2(\R^6)} \\ 
    	+ \meanlrlr{\tilde{\A}_{\p}}{-h^{2}\Delta_{\xi} + V(\xi/h) + E_{0}}{\tilde{\A}_{\p}}_{L^2(\R^6)}	\\
    	= \meanlrlr{\tilde{\A}_{\p}}{-\tfrac{1}{4}h^{2}\Delta_{\eta} + h^{2}W(\eta+\xi/2) - h^{2}D}{\tilde{\A}_{\p}}_{L^2(\R^6)}.
    }
   {where we used that $\alpha_{0}$ is the normalized zero energy eigenvector of the operator $-\Delta + V + E_{0}$.} The result then follows from next \cref{lemma: expectation W}.
    
%
    
	    In order to prove the second estimate, we use the cyclicity of the trace and the symmetry of the Laplacian, to get
    \bml{
       \tr \Delta \A_{\p}\overline{\A_{\p}}\A_{\p}\overline{\A_{\p}} = \braketr{\A_{\p}\overline{\A_{\p}}\A_{\p}}{ \tfrac{1}{2}(\Delta_{x}+\Delta_{y})\alpha_{\p}}_{L^{2}(\R^6)}	\\
       = \braketr{\tilde{\omega}_\psi}{\tfrac{1}{4}\Delta_{\eta} \tilde{\A}_{\p}}_{L^{2}(\R^6)} + \braketr{\tilde{\omega}_\psi}{\Delta_{\xi} \tilde{\A}_{\p}}_{L^{2}(\R^6)},
    }
    where we have set for short $ \tilde{\omega}_\psi(\eta,\xi) : = (\A_{\p}\overline{\A_{\p}}\A_{\p})(x,y) $.  Introducing the coordinates
    \begin{equation}
    \label{coordinates:n=4}
        X=\tfrac{1}{4}(x_{1}+x_{2}+x_{3}+x_{4}), \qquad \xi_{k}=x_{k+1}-x_{k}, \quad k=1,2,3,
    \end{equation}
 	and rescaling the relative ones, we obtain 
   \bml{
   \label{equation:traceCC}
        \tr (-h^{2}\Delta + E_{0}) \A_{\p}\overline{\A_{\p}} \A_{\p}\overline{\A_{\p}} =h\int_{\mathbb{R}^{12}}dX d\xi_{1} d\xi_{2} d\xi_{3} \: \p(X-hs) \overline{\p(X-ht)} \times \\
        \p(X+hs)\overline{\p(X+ht)} \lf[ \lf(-\Delta + E_{0} \ri) \Ao \ri] (\xi_{1}) \overline{\Ao(\xi_{2})}\Ao(\xi_{3})\overline{\Ao(\xi_{*})}	\\
        - \tfrac{1}{4} h^2 \braketr{\tilde{\omega}_\psi}{\Delta_{\eta} \tilde{\A}_{\p}}_{L^{2}(\R^6)},
   }
    where $\xi_{*}\vcentcolon=-\xi_{1}-\xi_{2}-\xi_{3}$ and $s,t$ are functions of $\xi_{1}, \xi_{2}, \xi_{3}$, i.e., 
    \begin{equation}
       s :=\tfrac{1}{4} \lf(\xi_{1}+2\xi_{2}+\xi_{3}\ri), \qquad t := \tfrac{1}{4}\lf( \xi_{3}-\xi_{1} \ri).
   \end{equation} 
    From this expression we are going to extract the quartic term needed to reconstruct the $\GP$ functional times $ h $ plus higher order contributions. The fundamental theorem of calculus allows to rewrite the first term on the r.h.s. of \eqref{equation:traceCC} as
   \bml{
    		\label{eqp: definition I1}
	       h\lVert\p\rVert^{4}_{L^{4}} \int_{\R^{9}}d\xi_{1} d\xi_{2} d\xi_{3} \:  \lf[ \lf(-\Delta + E_{0} \ri) \Ao (\xi_{1}) \ri] \overline{\Ao(\xi_{2})}\Ao(\xi_{3})\overline{\Ao(\xi_{*})} 	\\ 
	        + h\int_{\R^{12}} dX d\xi_{1} d\xi_{2} d\xi_{3} \int_{0}^{1}  d\tau \: \frac{d}{d\tau}\lf( \p(X-\tau hs) \overline{\p(X-\tau ht)} \times \ri. \\
	        \lf. \times \p(X+\tau hs)\overline{\p(X+\tau ht)} \ri) \lf[ \lf(-\Delta + E_{0} \ri) \Ao (\xi_{1}) \ri]\overline{\Ao(\xi_{2})}\Ao(\xi_{3})\overline{\Ao(\xi_{*})}	\\
	        = : h g_{\BCS} \lVert\p\rVert^{4}_{L^{4}} + h^2 I_1,
    }
    thanks to the explicit computation
    \bmln{
    		\int_{\R^{9}}d\xi_{1} d\xi_{2} d\xi_{3} \:  \lf[ \lf(-\Delta + E_{0} \ri) \Ao (\xi_{1}) \ri] \overline{\Ao(\xi_{2})}\Ao(\xi_{3})\overline{\Ao((\xi_{*}))}  \\
    		= (2\pi)^{3} \int_{\R^{3}} dp \: (p^{2}+E_{0}) \lvert \hat{\alpha}_0(p)\rvert^{4}.
   }
   Hence, \eqref{equation:traceCC} yields
   \beq
   		  \tr (-h^{2}\Delta + E_{0}) \A_{\p}\overline{\A_{\p}} \A_{\p}\overline{\A_{\p}} = h g_{\BCS} \lVert\p\rVert^{4}_{L^{4}} + h^2 \lf(I_1+ I_2\ri),
	\eeq
	where 
	\beq
		\label{eqp: definition I2}
		I_2 : = - \tfrac{1}{4} \braketr{\A_{\p}\overline{\A_{\p}}\A_{\p}}{\Delta_{\eta} {\A}_{\p}}_{L^{2}(\R^6)}.
	\eeq 
	The estimate on the term containing the external potential immediately follows from \cref{lemma: expectation W} using H\"{o}lder inequality with exponents $ \frac{1}{2}, \frac{1}{3} $ and $ \frac{1}{6} $:
    \bml{
		         \lf| \tr W \A_{\p}\overline{\A_{\p}}\A_{\p}\overline{\A_{\p}} \ri| \leq \tr \lf| W^{1/2} \A_{\p}\overline{\A_{\p}}\A_{\p}\overline{\A_{\p}} W^{1/2} \ri|	\\
		         \leq  \lf\lVert W^{1/2} \alpha_{\p} \ri\rVert_{\mathfrak{S}^{2}} \lf\lVert W^{1/2} \alpha_{\p} \ri\rVert_{\mathfrak{S}^{6}} \lf\lVert \overline{\A_{\p}} \alpha_{\p}\ri\rVert_{\mathfrak{S}^{3}} \leq \lf\lVert W^{1/2} \alpha_{\p} \ri\rVert^2_{\mathfrak{S}^{2}}  \lf\lVert  \alpha_{\p}\ri\rVert_{\mathfrak{S}^{6}}^2	\\
		          \leq C h \lVert \p \rVert_{L^{6}}^{2} \lf\lVert \hat{\alpha}_{0} \ri\rVert_{L^{6}}^{2} \lf\lVert W^{1/2} \alpha_{\p} \ri\rVert_{L^{2}}^{2} \leq C \lVert \p \rVert_{L^{6}}^{2}  \lf[ \lf\| W |\psi|^2 \ri\|_{L^1} + h A_0 \ri],
    }
   by the monotonicity of Schatten norms and \cref{proposition:estimates on alpha}.  The replacement of $ \lVert \p \rVert_{L^{6}}^2 $ with $ \lf\| \nabla \psi \ri\|_{L^2}^2 + \lf\| \psi \ri\|_{L^2}^2 $ {can be} done via Sobolev inequality.
\end{proof}

	\begin{lemma}
		\label{lemma: expectation W}
		Let $ \alpha_{\psi} $ be as \eqref{equation:alpha_p} and $ \mathrm{A}_0 $ as in \eqref{eq: A0}, then
		\beq
			\lf| \meanlrlr{\alpha_\psi}{W}{\alpha_\psi}_{L^2(\R^6)} - h^{-1} \int_{\R^{3}}d\eta \: W(\eta) \lvert \p\rvert^{2} \ri|\leq \mathrm{A}_0.
		\eeq
	\end{lemma}	
	
	\begin{proof}
		Using center-of-mass and relative coordinates as before, we get by the Taylor expansion \eqref{Taylor} 
		\bmln{
			\tfrac{1}{2}\meanlrlr{\alpha_\psi}{W(x) + W(y)}{\alpha_\psi}_{L^2(\R^6)} = \tfrac{1}{2}\meanlrlr{\tilde{\A}_{\p}}{{W(\eta+\xi/2)}}{\tilde{\A}_{\p}}_{L^2(\R^6)} \\ + \tfrac{1}{2}\meanlrlr{\tilde{\A}_{\p}}{{W(\eta-\xi/2)}}{\tilde{\A}_{\p}}_{L^2(\R^6)} \\
			= h^{-1} \int_{\R^{3}}d\eta \: W(\eta) \lvert \p(\eta)\rvert^{2}  
        + \tfrac{1}{2} h^{-4} \int_{\R^{6}}d\eta d\xi \: \xi\cdot \nabla W(\zeta) \lf| \psi(\eta) \ri|^2 \lf| \Ao(\xi/h) \ri|^2,
		}
		where we recall that $ \tilde{\alpha}_{\psi}(\eta, \xi) = \alpha_{\psi}(x,y) $. Hence, we have only to estimate the last term on the r.h.s. of the expression above: by {\cref{assumption:W} on $ W $, we deduce that, since $ \zeta \in (\eta, \eta + \xi/2) $,}
		 \bml{
    			\label{eqp: bound W}
       		\tfrac{1}{2} h^{-4}\int_{\R^{6}}d\eta d\xi \: \xi\cdot \nabla W(\zeta) \lf| \psi(\eta) \ri|^2 \lf| \Ao(\xi/h) \ri|^2 \\
       \leq C \int_{\R^{6}}d\eta d\xi \: \lf| \xi \ri| \lf( h^{\beta - 1} \lf|\xi \ri|^{\beta-1} + \lf| \eta \ri|^{\beta-1} + 1 \ri) \lf| \psi(\eta) \ri|^2 \lf| \Ao(\xi) \ri|^2
       } 
       which immediately implies the result, via the trivial bounds {$ |x|^{\beta-1} \leq W(x) + 1 $ (again by \cref{assumption:W})} and
       \beq
       	\lf\lVert \lvert\cdot \rvert^{\beta/2}\Ao \ri\rVert^{2}_{L^{2}} \leq C,	\qquad		\lf\lVert \lvert\cdot \rvert^{1/2}\Ao \ri\rVert^{2}_{L^{2}} \leq C,
	\eeq
	which follows from \eqref{property:decay}. 
		\end{proof}
		
 \begin{lemma}
    \label{lemma:estimatesUB}
    Let $\mathrm{I}_{1},\mathrm{I}_{2}$ as in \eqref{eqp: definition I1} and \eqref{eqp: definition I2}. Then, as $ h \to 0 $,$ \exists C < + \infty $ such that
   \beq
	      \lvert \mathrm{I}_{1}\rvert +  \lvert \mathrm{I}_{2}\rvert \leq  C \lVert \nabla\p \rVert_{L^{2}}^{4}.
    \eeq
    \end{lemma}

	\begin{proof}
		By \cite[Proof of Lemma 1]{BHS:paper}, there exist two finite constants $ C_{1}, C_2 $ such that
		\beqn
	       \lvert \mathrm{I}_{1}\rvert &\leq& C_{1} \lVert \nabla\p \rVert_{L^{2}}^{4} \lf\lVert \lvert \cdot \rvert\Ao \ri\rVert_{L^{2}} \lf\lVert \Ao \ri\rVert_{L^{2}} \lf\lVert \Ao \ri\rVert_{L^{1}} \lf\lVert V\Ao \ri\rVert_{L^{1}},	\nonumber \\
        \lvert \mathrm{I}_{2}\rvert &\leq& C_{2}\lVert \nabla\p\rVert^{4}_{L^{2}}. \nonumber
   		 \eeqn
   		 The result then follows from the properties of $ \alpha_0 $ {(see \cref{rem: alpha0})}.
	\end{proof}

\subsection{Energy upper bound}
\label{section:UB}
%
%

\label{subsection:}
{The result is obtained by testing the BCS energy functional on a suitable trial state.}
We define an admissible state $\Gamma_{\p}$, with off-diagonal element given by $ \alpha_{\psi} $ as in \eqref{equation:alpha_p} and $ \psi \in \mathscr{D}^{\GP} $, and upper left entry 
\beq
	\gamma_{\p} \vcentcolon = \Ap\overline{\Ap}+(1+\lambda h)\Ap\overline{\Ap}\Ap\overline{\Ap},
\eeq
for some $ \lambda \in \R^+ $.

\begin{remark}[Admissibility]
\label{rmk:on admissibility of trial state}
The admissibility requirement makes the correction of order $ \lambda h $ necessary. In fact, any correction of order $h^{\beta}$, $0 < \beta \leq 1$, {would work,} if $ \lambda $ is chosen appropriately, but $ \beta = 1 $ gives the best error bound {in our estimates}. Indeed,  the state is admissible if and only if $ \gamma - \gamma^2 - \alpha \overline{\alpha} \geq 0 $ {(see, e.g., \cite[Eq. (4.8)]{BHS:paper})}, which, assuming that the quartic correction is proportional to $ \lambda h^{\beta} $, yields the condition
    \begin{equation}
    \label{equation:equivalent adm}
    \lambda h^{\beta}-(1+\lambda h^{\beta})^{2}(\A_{\p}\overline{\A_{\p}})^{2}-2(1+\lambda h^{\beta})\A_{\p}\overline{\A_{\p}} \geq 0. 
    \end{equation}
    Since $\lVert \Ap \rVert_{\infty} \leq \lVert \Ap \rVert_{6} \leq C h^{1/2}$, this bound implies that we may choose $0<\beta <1$, and the latter condition would be satisfied for any value of $\lambda$. For $ \beta = 1 $, on the other hand, one is forced to take the parameter $\lambda$ large enough, but the inequality may still hold.
\end{remark}

We now apply \cref{prop:semiclassics for ub} to get
\bml{
    \mathcal{E}^{\BCS}_{\mu}[\Gamma_{\p}] = \tg_{\p} + \int_{\mathbb{R}^{6}} dx dy V\lf(\tfrac{x-y}{h}\ri)\lvert\alpha_{\p}(x,y)\rvert^2 \\
     = \tr\h\Ap\overline{\Ap} + \int_{\mathbb{R}^{6}} dx dy V\lf(\tfrac{x-y}{h}\ri)\lvert\alpha_{\p}(x,y)\rvert^2 + (1+\lambda h )\tr \h\Ap \overline{\Ap}\Ap \overline{\Ap} \\
      \leq h \int_{\R^{3}} d\eta \lf\{ \tfrac{1}{4} \lvert \nabla\p \rvert^{2}+(W-D)\lvert \p \rvert^{2} + g_{\BCS} \lvert \p\rvert^{4} \ri\} \\
      + C h^{2} \lf[  \lVert \nabla\p\rVert^{4}_{L^{2}} + \lf\lVert W \lvert\p\rvert^{2} \ri\rVert_{L^{1}}^2  + \lf\| \psi \ri\|_{L^2}^4 + 1 \ri]
}
as $h\rightarrow 0$. The upper bound \eqref{eq: upper bound} is thus a straightforward consequence of \eqref{eq: a priori psi}.
\subsection{Energy lower bound}
\label{section:LB}

We consider any admissible $\BCS$ state $\Gamma$ satisfying $\EBCS\leq C_{\Gamma} h$, whose existence is ensured by the analysis in the previous \cref{section:UB}. The integral kernel of $\alpha$, the upper-right entry of $\Gamma$, can be {decomposed} as
    \begin{equation}
    		\label{eq: decomposition}
        \alpha(x,y) = \alpha_{\psi}(x,y) + r(x,y) = h^{-2}\p\lf( \tfrac{x+y}{2} \ri)\Ao\lf( \tfrac{x-y}{h} \ri)+r(x,y).
    \end{equation}
    where $ r $ is chosen to be orthogonal to $ \alpha_0 $:
    \beq
    		\label{eq: orthogonality}
    		\braketr{\alpha_0\lf( \, \cdot \,/h \ri)}{\tilde{r}}_{L^2_\xi(\R^3)} = 0,
	\eeq
	where $ \tilde{r}(\eta,\xi) : = r(x,y) $ and the coordinates $ \eta, \xi $ are defined in \eqref{eq: center of mass}. With such a choice, the order parameter $ \psi $ is naturally defined in terms of  $\alpha $ as (recall the notation $ \tilde{\alpha}(\eta,\xi) : = \alpha(x,y) $)
	\begin{equation}
    			\p(\eta)\vcentcolon= h^{-1}\braketr{\alpha_0\lf( \, \cdot \,/h \ri)}{\tilde{\alpha}}_{L^2_{\xi}(\R^3)} =  h^{-1}\int_{\R^{3}}d\xi \: \Ao(\xi/h) \tilde{\alpha}(\eta,\xi),
	\end{equation}
	Note also that, because of the orthogonality of $ r $ to $ \alpha_0 $, one immediately gets
	\beq
		\lVert \alpha \rVert^{2}_{L^{2}(\R^{6})} = \lVert \alpha_{\p}\rVert^{2}_{L^{2}(\R^{6})} + \lVert r\rVert_{L^{2}(\R^{6})}^{2}  = h^{-1}\lVert \p\rVert^{2}_{L^{2}} + \lVert r\rVert^{2}_{L^{2}}.
	\eeq

The physical meaning of such a decomposition is apparent: $ \alpha $ represents the wave function of a pair of particles and it almost factorizes in the coordinates of the center-of-mass reference frame. More precisely, $\alpha_{0}$ describes the wave function in the relative coordinate living on the microscopic scale $h$, while $\p$ is the wave function in the in center-of-mass coordinate and varies on the macroscopic scale.

%
%
%
%

We start with a preliminary lower bound on the $\BCS$ energy functional in terms of the off diagonal entry $\alpha$ of $ \Gamma $. Indeed, for any admissible $ \Gamma $, it can be seen that one can bound $ \EBCS $ from below in terms of a functional of $ \alpha $ alone. 

\begin{lemma}
	\label{lemma: reduction alpha}
	Let $\mu = -E_{0}+h^{2}D$, $ D\in \R$. For any admissible $ \Gamma $ and for $ h $ small enough,
	\beq
		\label{eq: alpha energy}
		\EBCS \geq \tr \h \alpha \overline{\alpha} + \tr \h \alpha \overline{\alpha} \alpha \overline{\alpha} + \int_{\mathbb{R}^{6}}dx dy \: V\lf(\tfrac{x-y}{h}\ri)\lvert\alpha(x,y)\rvert^2.
	\eeq
\end{lemma}

\begin{proof}
	The proof is given, e.g., in \cite[Proposition 6.2]{FLS:paper}. We spell it in details here for the sake of {completeness}. The admissibility of $ \Gamma $, i.e., the condition $ 0 \leq \Gamma \leq 1 $, is equivalent to 
	\beq
		\label{eq: admissible}
		\gamma - \gamma^2 - \alpha \overline{\alpha} \geq 0.
	\eeq
	Since for $ h $ small enough $ \h $ is positive, as it follows from the trivial bound
	\beq
		\label{eq: h lower bound}
		\h \geq E_0 - D h^2 > 0,
	\eeq
	we can use the monotonicity of the trace and apply the above inequality to get the result, since \eqref{eq: admissible} implies that $ \gamma \geq \alpha \overline{\alpha} + \alpha \overline{\alpha}\alpha \overline{\alpha} $ (see \cite[Eq. (6.2)]{FLS:paper}).
\end{proof}

The next lower bound give more information on the decomposition \eqref{eq: decomposition}.

\begin{lemma}
\label{proposition:mixedterms}
    Let $\mu=-E_{0}+h^{2}D$, $D\in \R$, and let $\Gamma$ an admissible $\BCS$ state with upper-right entry $\alpha$ as in \eqref{eq: decomposition}. Then, there exists a finite constant $ C $ such that (recall \eqref{eq: A0}), as $ h \to 0 $,
    \bml{
        \tr \h \alpha \overline{\alpha} + \int_{\mathbb{R}^{6}}dx dy \: V\lf(\tfrac{x-y}{h}\ri)\lvert\alpha(x,y)\rvert^2 \geq  h \meanlrlr{\psi}{- \tfrac{1}{4} \Delta +  W - D}{\psi}_{L^2(\R^3)} \\
        + \tfrac{1}{2} g \lVert r \rVert^{2}_{L^{2}(\R^6)}
        + h^{2} \lf[ \meanlrlr{\tilde{r}}{- \tfrac{1}{4} \Delta_\eta  - \varepsilon \Delta_\xi + \tfrac{1}{2} W - D}{\tilde{r}}_{L^2(\R^6} - A_0  \ri].
    \label{equation:apriori1}
    }
\end{lemma}

\begin{proof}
Plugging in the operator bound \eqref{eq: h lower bound}, we can immediately get rid of the second term in \eqref{eq: alpha energy} to obtain
\bdm
	\EBCS \geq \tr \h \alpha \overline{\alpha} + \int_{\mathbb{R}^{6}}dx dy \: V\lf(\tfrac{x-y}{h}\ri)\lvert\alpha(x,y)\rvert^2 + (E_{0}-h^{2}D)\lVert \alpha\rVert_{\mathfrak{S}^{4}}^{4},
\edm
and the last term can be dropped since it is positive. Next, we estimate the first term, which reads 
\bmln{
    \tr \h \alpha \overline{\alpha} + \int_{\mathbb{R}^{6}}dx dy \: V\lf(\tfrac{x-y}{h}\ri)\lvert\alpha(x,y)\rvert^2 \\
    = \int_{\R^{6}} d \eta d \xi \: \overline{\tilde{\alpha}}(\eta,\xi) \lf( -\tfrac{1}{4} h^2 \Delta_{\eta}- h^{2}\Delta_{\xi} + h^{2}W(\eta+\xi/2) + \ri.	\\
    \lf. V(\xi/h) - \mu \ri) \Tilde{\alpha}(\eta,\xi).
}
By plugging in the decomposition \eqref{eq: decomposition}, we get
\bml{
	 \tr \h \alpha \overline{\alpha} = \meanlrlr{\alpha_{\psi}}{\h}{\alpha_{\psi}} + \int_{\mathbb{R}^{6}}dx dy \: V\lf(\tfrac{x-y}{h}\ri)\lvert\alpha_{\psi}(x,y)\rvert^2 \\
	 + \meanlrlr{r}{\h}{r} + \int_{\mathbb{R}^{6}}dx dy \: V\lf(\tfrac{x-y}{h}\ri)\lvert r(x,y)\rvert^2 + 2 h^2 \Re \braketr{\alpha_{\psi}}{W r},
}
since the potential $ W $ is the only operator which does not factorize in the decomposition $ L^2(\R^6) = L^2_{\eta}(\R^3) \otimes L^2_{\xi}(\R^3) $. The sum of the first two terms has already been estimated in \cref{prop:semiclassics for ub}, so that it just remains to consider the quadratic expression on $r $ and the mixed term.

The mixed term can be controlled by exploiting the Taylor expansion \eqref{Taylor} and the orthogonality \eqref{eq: orthogonality}, obtaining
\bmln{
        2 h^2 \lf| \Re \braketr{\alpha_{\psi}}{W r} \ri| = 2 \lf| \int_{\R^6} d \eta d \xi \: \xi \cdot \nabla W( \zeta) \overline{\psi}(\eta) \alpha_{0}(\xi/h) \tilde{r}(\eta,\xi) \ri| \\
        \leq C \int_{\R^6} d \eta d \xi \: \lf| \xi \ri| \lf( \lf| \xi \ri|^{\beta-1} + \lf| \eta \ri|^{\beta} + 1 \ri) |\psi(\eta)| |\alpha_0(\xi/h)| \lf| \tilde{r}(\eta, \xi) \ri|  
	}
	by the trivial bound $ |\eta|^{\beta-1} \leq |\eta|^{\beta} + 1 $. Hence, by Cauchy-Schwarz inequality we get
	\bmln{
		2 h^2 \lf| \Re \braketr{\alpha_{\psi}}{W r} \ri| \leq C \lf\| \psi \ri\|_{L^2(\R^3)} \lf\| r \ri\|_{L^2(\R^6)}  \times	\\
		\times
		\lf( \int_{\R^3} d \xi \: \lf( |\xi|^{2\beta} + \lf| \xi \ri|^2 \ri) \lf| \alpha_0(\xi/h) \ri|^2 \ri)^{1/2} \\
		+ C \lf( \lf\| W |\psi|^2 \ri\|_{L^1(\R^3)}^{1/2} + \lf\| \psi \ri\|_{L^2(\R^3)} \ri) \lf[ \lf( \int_{\R^6} d \eta d \xi \: |\eta|^{\beta} \lf| \tilde{r} \ri|^2 \ri)^{1/2} + \lf\| r \ri\|_{L^2(\R^6)} \ri] \times	\\
		\times \lf( \int_{\R^3} d \xi \: |\xi|^2 \lf| \alpha_0(\xi/h) \ri|^2 \ri)^{1/2}	\\
		\leq C h^{5/2} \lf( \lf\| \psi \ri\|^2_{L^2} + \lf\| r \ri\|^2_{L^2} + \lf\| W |\psi|^2 \ri\|_{L^1} + \lf\| W |r|^2 \ri\|_{L^1} \ri)
	}
	where we have estimated 
	\bdm
		\int_{\R^6} d \eta d \xi \: |\eta|^{\beta} \lf| \tilde{r} \ri|^2 \leq C \lf\| W \lf| r \ri|^2 \ri\|_{L^1(\R^6)}.
	\edm
	The two terms depending on $ r $ can then be absorbed in the corresponding positive ones coming from the estimate of $ \meanlrlr{r}{\h}{r} $ by adding a $ \frac{1}{2}$  prefactor for $ h$  small enough, while the other two can be included in the $ A_0 h^2 $ remainder up to the change of the constant $ C $ in $ A_0 $.
	
  	The quadratic expression in $ \xi $ is bounded from below by means of \cref{hypothesis:spectralgap}:
	    \bml{
	    	\int_{\R^{3}} d\eta \: \meanlrlr{r(\eta,\cdot)}{-h^{2}\Delta_{{\xi}} +V(\, \cdot \,/h)+E_{0}}{r(\eta,\cdot)}_{L^2_\xi(\R^3)}  \\  
	    	\geq \int_{\R^{3}} d\eta \: \meanlrlr{r(\eta,\cdot)}{-h^{2} \varepsilon \Delta_{{\xi}} + g}{r(\eta,\cdot)}_{L^2_\xi(\R^3)}	\\
	    	= g\lVert r\rVert_{L^{2}({\R^{6}})}^{2} + h^{2}\varepsilon \lf\| \nabla_{\xi} r \ri\|_{L^2(\R^6)}^2.
		}
		\end{proof}

\begin{lemma}
	\label{lemma: quartic}
	 Let $\mu=-E_{0}+h^{2}D$, $D\in \R$, and let $\Gamma$ an admissible $\BCS$ state with upper-right entry $\alpha$ as in \eqref{eq: decomposition}, such that $ \EBCS \leq C_{\Gamma} h $. Then, there exists a finite constant $ C $ such that
    \beq
        \lf| \tr \h \alpha \overline{\alpha} \alpha \overline{\alpha} - \tr \h \alpha_{\psi} \overline{\alpha}_{\psi} \alpha_{\psi} \overline{\alpha}_{\psi} \ri| \leq  C h^{2} \lf( \lf\| \nabla \psi \ri\|^4_{L^2} + {A_0^2}  \ri).
    \label{eq: quartic below}
    \eeq
\end{lemma}

\begin{proof}
	We first rewrite the quartic term via
	\bdm
		\alpha\overline{\alpha}\alpha\overline{\alpha} - \alpha_{\p}\overline{\alpha_{\p}}\alpha_{\p}\overline{\alpha_{\p}} = r\overline{\alpha}\alpha\overline{\alpha_{\p}}+\alpha_{\p}\overline{\alpha}\alpha\overline{r}+r\overline{\alpha}\alpha\overline{r} + \alpha_{\p}(\overline{\alpha}\alpha-\overline{\alpha_{\p}}\alpha_{\p})\overline{\alpha_{\p}},
	\edm
so that the cyclicity of trace and triangle inequality yields
	\bml{
			\label{eqp: 76}
			\lf| \tr \mathfrak{h}\alpha\overline{\alpha}\alpha\overline{\alpha} - \tr \mathfrak{h}\alpha_{\p}\overline{\alpha_{\p}}\alpha_{\p}\overline{\alpha_{\p}} \ri| \leq \lf\|  \mathfrak{h}^{1/2}r\overline{\alpha}\alpha\overline{\alpha_{\p}}  \mathfrak{h}^{1/2}\ri\|_{\mathfrak{S}^{1}} +  \lf\| \mathfrak{h}^{1/2}\alpha_{\p}\overline{\alpha}\alpha\overline{r} \mathfrak{h}^{1/2}\ri\|_{\mathfrak{S}^{1}}\\
			+\lf\| \mathfrak{h}^{1/2} r\overline{\alpha}\alpha\overline{r}\mathfrak{h}^{1/2}\ri\|_{\mathfrak{S}^{1}} + \lf\| \mathfrak{h}^{1/2}\alpha_{\p}(\overline{\alpha}\alpha-\overline{\alpha_{\p}}\alpha_{\p})\overline{\alpha_{\p}} \mathfrak{h}^{1/2} \ri\|_{\mathfrak{S}^{1}}.
	}
	To estimate this four terms, we apply H\"{o}lder inequality:
	\beqn
   		\lf\| \mathfrak{h}^{1/2}\alpha_{\p}\overline{\alpha}\alpha\overline{r} \mathfrak{h}^{1/2}\ri\|_{\mathfrak{S}^{1}} 	&\leq&  \lf\| \mathfrak{h}^{1/2}\alpha_{\p}\ri\|_{\mathfrak{S}^{6}}\lf\| \alpha\ri\|_{\mathfrak{S}^{6}}^{2}\lf\|\mathfrak{h}^{1/2} r\ri\|_{\mathfrak{S}^{2}};	\nonumber \\
   		\lf\|\mathfrak{h}^{1/2}r\overline{\alpha}\alpha\overline{\alpha_{\p}}\mathfrak{h}^{1/2}\ri\|_{\mathfrak{S}^{1}} &\leq&  \lf\| \mathfrak{h}^{1/2}r \ri\|_{\mathfrak{S}^{2}} \lf\|\alpha\ri\|_{\mathfrak{S}^{6}}^{2} \lf\|\mathfrak{h}^{1/2}\alpha_{\p}\ri\|_{\mathfrak{S}^{6}};	\nonumber \\
   		\lf\|\mathfrak{h}^{1/2} r\overline{\alpha}\alpha\overline{r}\mathfrak{h}^{1/2}\ri\|_{\mathfrak{S}^{1}} = \lf\|\alpha\overline{r}\mathfrak{h}^{1/2}\ri\|_{\mathfrak{S}^{2}}^{2} &\leq& \lf\|\alpha\ri\|_{\mathfrak{S}^{\infty}}^{2}\lf\|\mathfrak{h}^{1/2} r\ri\|_{\mathfrak{S}^{2}}^{2};	\nonumber \\
   		\lf\|\mathfrak{h}^{1/2}\alpha_{\p}(\alpha\overline{\alpha}-\overline{\alpha_{\p}}\alpha_{\p})\overline{\alpha_{\p}}\mathfrak{h}^{1/2} \ri\|_{\mathfrak{S}^{1}} &\leq& \lf\|\alpha\overline{\alpha}-\overline{\alpha_{\p}}\alpha_{\p}\ri\|_{\mathfrak{S}^{3/2}} \lf\|\mathfrak{h}^{1/2}\alpha_{\p} \ri\|_{\mathfrak{S}^{6}}^{2}. \nonumber
	\eeqn
	Plugging the above bounds in \eqref{eqp: 76}, we obtain
	\bml{
		 \lf| \tr \mathfrak{h}\alpha\overline{\alpha}\alpha\overline{\alpha} - \tr \mathfrak{h} \alpha_{\p}\overline{\alpha_{\p}}\alpha_{\p} \overline{\alpha_{\p}} \ri| \leq 2\lf\| \mathfrak{h}^{1/2}\alpha_{\p}\ri\|_{\mathfrak{S}^{6}}\lf\| \alpha\ri\|_{\mathfrak{S}^{6}}^{2}\lf\|\mathfrak{h}^{1/2}  r\ri\|_{\mathfrak{S}^{2}} \\	
		 + \lf\|\alpha\ri\|_{\mathfrak{S}^{{6}}}^{2}\lf\|\mathfrak{h}^{1/2} r\ri\|_{\mathfrak{S}^{2}}^{2}+\lf\|\overline{\alpha}\alpha-\overline{\alpha_{\p}}\alpha_{\p}\ri\|_{\mathfrak{S}^{3/2}} \lf\|\mathfrak{h}^{1/2}\alpha_{\p} \ri\|_{\mathfrak{S}^{6}}^{2}.
	}
	
	By \eqref{equation:apriori1} and the condition on the BCS energy of $ \Gamma $, we deduce the inequality
	\bml{
		\label{eq: first bound r}
		  \lf(  \tfrac{1}{2} g - D h^2 \ri) \lf\| r \ri\|^{2}_{L^{2}} + h^{2} \lf[ \meanlrlr{\tilde{r}}{- \tfrac{1}{4} \Delta_\eta  - \varepsilon \Delta_\xi + {\tfrac{1}{2}} W}{\tilde{r}}_{L^2(\R^6)} \ri] \\
		  \leq  C h \lf[ 1 + \lf\| \psi \ri\|_{L^2}^2 +   h {A_0}  \ri],
	}
	{with $A_{0}$ defined in \cref{prop:semiclassics for ub}}. This, for $ h $ small enough (e.g., smaller than $ \sqrt{g/{(4D)}} $), gives a bound on $ \lf\| r \ri\|^{2}_{L^{2}} $ as well as its Sobolev norms in terms of the norm of $ \psi $.
	Hence, we have
		\bmln{
			\lf\|\overline{\alpha}\alpha-\overline{\alpha_{\p}}\alpha_{\p}\ri\|_{\mathfrak{S}^{3/2}} = \lf\|\overline{\alpha_{\p}}r + \overline{r} \alpha_{\p}+\overline{r}r \ri\|_{\mathfrak{S}^{3/2}}\\
				\leq  2 \lf\|\Ap \ri\|_{\mathfrak{S}^{6}}\lf\| r \ri\|_{\mathfrak{S}^{2}} +
				\lf\| r \ri\|^{{2}}_{\mathfrak{S}^{2}}	
				 \leq C h \lf[ \lf\| \psi \ri\|^{{2}}_{L^6} + \lf\| \psi \ri\|^{{2}}_{L^2} + 1 +  {h A_0} \ri],
	}
	by the monotonicity of Schatten norms, \cref{proposition:estimates on alpha} and \eqref{eq: first bound r}. Similarly, {by Sobolev inequality}
\bml{
    \lf\| \alpha \ri\|_{\mathfrak{S}^{6}}^{2}\leq C \lf( \lf\| \Ap \ri\|_{\mathfrak{S}^{6}}^{2} + \lf\| r \ri\|^2_{\mathfrak{S}^{2}} \ri) \leq C h \lf[ \lf\| \psi \ri\|^2_{L^6} +  \lf\| \psi \ri\|_{L^2}^2 + 1+  h {A_0} \ri]	\\
    {\leq C h \lf[ \lf\| \nabla \psi \ri\|^2_{L^2} +  \lf\| \psi \ri\|_{L^2}^2 + 1 +  h {A_0} \ri].}
}
	To conclude, we have to estimate the norms of $ \h^{1/2} \alpha_{\psi} $ but, for any operator $T$, one has
	\bmln{
		    \lf\|  \h^{1/2}T \ri\|_{\mathfrak{S}^{2n}} = \lf\| T^{\dagger} \h T \ri\|_{\mathfrak{S}^{n}}^{1/2} 
		    \\  \leq \lf( h^2 \lf\| T^{\dagger}(- \Delta)T \ri\|_{\mathfrak{S}^{n}}+ h^{2}\lf\| T^{\dagger}W T \ri\|_{\mathfrak{S}^{n}} + \mu\lf\| T^{\dagger}T \ri\|_{\mathfrak{S}^{n}} \ri)^{1/2} \\ 
     \leq h \lf(\tfrac{1}{2}\lf\| \nabla_{\eta} T \ri\|_{\mathfrak{S}^{2n}} + \lf\| \nabla_{\xi} T \ri\|_{\mathfrak{S}^{2n}} + \lf\| W^{1/2}T\ri\|_{\mathfrak{S}^{2n}} \ri) + (E_{0}-h^{2}D)\lf\| T\ri\|_{\mathfrak{S}^{2n}}.
	}
	Applying this inequality to estimate the norms above and using once more the monotonicity of Schatten norms, {\cref{prop:semiclassics for ub}, \cref{lemma: expectation W,proposition:estimates on alpha} and Sobolev inequality}, 
	we obtain
	\bmln{
    		\lf\|  \mathfrak{h}^{1/2} \alpha_{\p} \ri\|_{\mathfrak{S}^{6}} \leq h \lf[ \tfrac{1}{2}\lf\|  \nabla_{\eta} \tilde{\alpha}_{\p} \ri\|_{\mathfrak{S}^{2}} + \lf\| \nabla_{\xi} \tilde\alpha_{\p} \ri\|_{\mathfrak{S}^{{2}}}+ \lf\| W^{1/2}\alpha_{\psi} \ri\|_{\mathfrak{S}^{2}} \ri] + E_{0} \lf\| \alpha_{\p} \ri\|_{\mathfrak{S}^{6}}\\
    		 \leq C h^{1/2} \lf[ \lf\| W |\psi|^2 \ri\|^{1/2}_{L^1} + \lf\| \psi \ri\|_{L^6} +  A_0 + {h \lf\| \nabla \psi \ri\|_{L^2}} \ri]	\\
    		 {\leq C h^{1/2} \lf[\lf\| \nabla \psi \ri\|_{L^2} + \lf\| W |\psi|^2 \ri\|^{1/2}_{L^1} +  A_0  \ri]},
	}
	\bmln{
    		\lf\|  \mathfrak{h}^{1/2} r \ri\|_{\mathfrak{S}^{2}} \leq  h \lf[ \tfrac{1}{2} \lf\|  \nabla_{\eta} r \ri\|_{\mathfrak{S}^{2}}+ \lf\| \nabla_{\xi} r \ri\|_{\mathfrak{S}^{2}}+ \lf\| W^{1/2}r\ri\|_{\mathfrak{S}^{2}} \ri] + E_{0} \lf\| r \ri\|_{\mathfrak{S}^{2}}\\
    		 \leq  C h^{1/2} \lf[ 1 + \lf\| \psi \ri\|_{L^2} +   h^{1/2} {\sqrt{A_0}} \ri],
	}
{as follows from the a priori estimate \eqref{eq: first bound r}}. 
Putting together all the bounds found so far, we get the result.
\end{proof}

In order to complete the proof of the lower bound, we need a last ingredient.

\begin{lemma}	
	\label{lemma: psi}
	Let $\mu=-E_{0}+h^{2}D$, $D\in \R$, and let $\Gamma$ an admissible $\BCS$ state with upper-right entry $\alpha$ as in \eqref{eq: decomposition}, such that $ \EBCS \leq C_{\Gamma} h $. Then, there exists a finite constant $ C $ such that
	\beq
		\label{eq: psi}
		\int_{\R^3} d \eta \: \lf\{ |\nabla \psi|^2 + W |\psi|^2 + |\psi|^2 + |\psi|^4 \ri\} \leq C.
	\eeq
\end{lemma}

\begin{proof}
	Let us denote for short 
	\bdm
		\E : = \int_{\R^3} d \eta \: \lf\{ |\nabla \psi|^2 + W |\psi|^2 + |\psi|^2 + |\psi|^4 \ri\}.
	\edm
	Combining \cref{lemma: quartic} with \eqref{equation:quartic}, we get
	\beq
		\tr \h \alpha \overline{\alpha} \alpha \overline{\alpha} \geq g_{\BCS} h \lf\| \psi \ri\|_{L^4}^4 - C h^{2} \E^2,
	\eeq
	so that, by \cref{proposition:mixedterms}, we find
	\beq
		\label{eqp: lower bound}
		C_{\Gamma} h \geq \EBCS \geq h \EGP - C h^{2} \lf( \E^2 + 1 \ri),
	\eeq
	where we used once more the estimate on $ \lf\| r \ri\|_{L^2} $ following from \eqref{eq: first bound r}. Since there exists a positive constant $ c > 0 $ such that $ \EGP \geq c \E - D \lf\| \psi \ri\|_{L^2}^2 $, we get
	\bdm
		\E \leq \tfrac{1}{c} \lf( C_{\gamma} + D \lf\| \psi \ri\|_{L^2}^2 \ri) + O(h).
	\edm
	However, such a bound gives a control on the norms $ \lf\| W |\psi|^2 \ri\|_{L^{1}} $ and $ \lf\| \psi \ri\|_{L^4} $, which can be used as in the proof of \cref{pro: a priori psi} to get an estimate of $ \lf\| \psi \ri\|_{L^2}^2 $, i.e., one obtains that there exists a finite constant such that
	\beq
		\lf\| \psi \ri\|_{L^2}^2 \leq C,
	\eeq
	which in turn yields the result.
\end{proof}

The estimate \eqref{eqp: lower bound} together with \eqref{eq: psi} gives the energy lower bound \eqref{eq: lower bound}. The combination of \cref{lemma: reduction alpha,proposition:mixedterms,lemma: quartic,lemma: psi} provides the proof of the remaining statements about the decomposition of $ \alpha $.

\end{document}